\documentclass{PoS}

\usepackage{graphicx}
\usepackage{mathrsfs}
\usepackage[authoryear,square]{natbib}
\bibpunct{(}{)}{;}{a}{}{,}

\title{Testing foundations of modern cosmology with SKA all-sky surveys}

\ShortTitle{Foundations of modern cosmology}

\author{\speaker{Dominik J. Schwarz},$^1$ David Bacon,$^2$ 
Song Chen,$^1$ Chris Clarkson,$^3$ Dragan Huterer,$^4$ Martin Kunz,$^{5,6}$ Roy Maartens,$^{7,2}$ 
Alvise Raccanelli,$^{8,9,10}$ Matthias Rubart,$^1$ Jean-Luc Starck$^{11}$ \\
        $^1$Fakult\"at f\"ur Physik, Universit\"at Bielefeld, 33501 Bielefeld, Germany; 
        $^2$Institute of Cosmology \& Gravitation, University of Portsmouth, Portsmouth 
        PO1 3FX, United Kingdom; 
        $^3$Centre for Astrophysics, Cosmology \& Gravitation and 
                Department of Mathematics \& Applied Mathematics, University of Cape Town, Cape Town 7701, 
                South Africa;
         $^4$Department of Physics, University of Michigan, Ann Arbor, MI 48109-1040, USA;        
         $^5$Universit\'e de Gen\`eve, D\'epartement de Physique Th\'eorique and CAP, CH-1211 Gen\`eve 4, Switzerland;
        $^6$African Institute for Mathematical Sciences, Cape Town 7945, South Africa;
        $^7$Physics Department, University of the Western Cape, Cape Town 7535, South Africa;
        $^8$Department of Physics \& Astronomy, Johns Hopkins University, 3400 N.~Charles St., 
                  Baltimore, MD 21218, USA
        $^9$Jet Propulsion Laboratory, California Institute of Technology, Pasadena CA 91109, USA;
        $^{10}$California Institute of Technology, Pasadena CA 91125, USA;
        $^{11}$Laboratoire AIM, UMR CEA-CNRS-Paris, Irfu, SAp, CEA Saclay, F-91191 Gif-sur-Yvette CEDEX,
              France \\
        E-mail: \email{dschwarz@physik.uni-bielefeld.de}}
        
\abstract{Continuum and HI surveys with the Square Kilometre Array (SKA) will allow us to probe 
some of the most fundamental assumptions of modern cosmology, including the Cosmological Principle. 
SKA all-sky surveys will map an enormous slice of space-time and reveal cosmology at 
superhorizon scales and redshifts of order unity. We illustrate the potential of these surveys
and discuss the prospects to measure the cosmic radio dipole at high fidelity. We outline several 
potentially transformational tests of cosmology to be carried out by means of SKA all-sky surveys.}

\FullConference{
Advancing Astrophysics with the Square Kilometre Array\\
June 8-13, 2014\\
Giardini Naxos, Italy}

\begin{document}

\section{Introduction}

The Square Kilometre Array (SKA) will allow us to test fundamental assumptions of modern 
cosmology at redshifts of order unity and at an accuracy level matching and complementing the high 
fidelity observations of the cosmic microwave background (CMB). 

The Cosmological Principle 
states that the Universe is spatially 
isotropic and homogeneous. 
This holds on sufficiently large scales and needs to be interpreted in a statistical sense. 
Historically, it provided a very powerful motivation
to single out the Friedmann-Lema\^itre models despite a lack 
of knowledge regarding the initial conditions of the Universe. Cosmological inflation, proposed 
in the 1980s, allowed the Universe to start from a reasonably 
small patch of almost homogeneous and isotropic space.  According to the idea of cosmological 
inflation, suddenly at least one small patch is inflated to contain today's observable Universe. In the 
course of inflation, any previously existing anisotropy or inhomogeneity is exponentially 
diluted. However, unavoidable quantum fluctuations are squeezed by the rapid expansion during 
inflation and become the seeds for large scale structure. 
The result is a statistically isotropic and homogeneous Universe (at least locally). 

These ideas are confirmed by the observed high degree of isotropy of CMB radiation, which enables 
us to define a CMB frame by measuring a temperature monopole, 
$T_0 = 2.2755 \pm 0.0006\,$K and dipole, $T_1 = 3.355 \pm 0.008\,$mK towards 
$(l,b) = (263.99^\circ \pm 0.14^\circ, 48.26^\circ \pm 0.03^\circ)$ \citep{Hinshaw}. 
The concept of a spatially homogeneous Universe allows us to speak about a cosmic time or 
an age of the Universe. By measuring the CMB temperature $T_0$ and the present expansion 
rate of the Universe $H_0$ we can anchor the thermal history of the Universe to its expansion history.

Radio surveys played an important role to establish that the Universe extends to redshifts 
beyond unity and that it is almost isotropic [see e.g. \cite{Ryle1961}]. Today,
observations of the CMB confirm the predictions of cosmological inflation impressively 
(Planck collaboration \citeyear{Planck2013XVI,Planck2013XXII}). 
However, it is unknown for how long (or how many e-foldings) inflation took place. 
In order to explain the observed spatial flatness of the Universe, about 50 to 60 e-foldings could be enough, 
but in many models it took much longer, i.e. the domain in which the statistical cosmological principle 
applies is expected to be much larger than the observable Universe.  The quest to determine the 
duration of inflation, as well as the related question of the topology of the Universe, 
can only be answered by observing the biggest scales.

Interestingly enough, the CMB exhibits unexpected features at the largest angular scales, 
among them a lack of angular correlation, alignments between the dipole, quadrupole and 
octupole, hemispherical asymmetry, a dipolar power modulation, and parity asymmetries 
(Planck collaboration \citeyear{Planck2013XXIII}; Copi et al. \citeyear{Copi:2013cya, Copi:2013jna}). 
Understanding the statistical significance of these anomalies is a hot topic 
\citep{WMAP2013,Starck2013,Rassat2014} since lack of statistical 
isotropy or Gaussianity could rule out the standard cosmological model.  
As the precision of these CMB measurements is limited by our understanding of the foregrounds and 
observational uncertainties are already much smaller than the cosmic variance at those scales,   
it is very hard to identify the cause of these anomalies without an independent probe at the same scales.

However, based on the observed CMB anisotropies and despite of these anomalies,
deviations from statistical isotropy have to be small. The observational situation is less clear for 
the case of statistical homogeneity, as testing the assumption of isotropy is much simpler than testing 
homogeneity \citep{Maartens:2011yx,Clarkson:2012bg}. 

SKA will probe an enormous number of independent modes when studying the large-scale 
structure of the Universe and will measure superhorizon sized modes at redshifts of order unity (better 
than any existing or planned infrared, optical, or X-ray campaign). This will enable us to probe scales 
that have not been in causal contact since the first horizon crossing during inflation and that contain 
information that was frozen in during cosmological inflation. In contrast to the CMB, the radio sky provides 
a probe of those largest scales at a redshift of order unity (2D for continuum surveys and 3D for HI surveys). 

SKA would enable several tests of the fundamental cosmological
principles. For example, the rest frames of the CMB and large scale structure (LSS) may not coincide
due to novel superhorizon physics --- for example, presence of isocurvature
modes \citep{erickcek:2008jp}. SKA's width and depth will enable a measurement
of the kinematic dipole with respect to the LSS reference frame via the
relativistic aberration and Doppler shift (``bunching up'' of SKA sources in the direction of
the dipole). This, when combined with the CMB's own measurement of the
kinematic dipole would, for the first time, enable the test of whether the two
reference frames --- that of  the CMB and the LSS --- are one and the same, as demanded by the Cosmological Principle.

Here we describe how to use 
all-sky ($3 \pi$) SKA continuum surveys to test statistical isotropy and to measure the cosmic dipole 
and other low-$\ell$ multipole moments. These issues are tightly connected to tests of non-Gaussianity 
and the topology of the Universe, the former aspect is described in Camera et al.~(\citeyear{Camera}).
All-sky 
SKA HI threshold surveys will additionally allow us to test the homogeneity of the Universe at 
superhorizon scales -- a test that has never before been performed. Statistical homogeneity and isotropy 
are assumed to hold true in other cosmology-related contributions to this book (Bull et al. \citeyear{BAO}; 
Raccanelli et al. \citeyear{RSD}).    
Tests of statistical isotropy and homogeneity will also allow (and force) us to dig deep into the systematics of 
SKA surveys and thus help to put all cosmological and non-cosmological results of SKA surveys 
on firm grounds. 

The conceptually simplest probe of cosmology is differential number counts \citep{deZotti}. If no redshift 
information is available, one can count the number of (extragalactic) radio sources per solid angle and 
flux density. Besides flux calibration issues, the cosmological information contained in differential 
number counts is limited by the diversity of radio sources and their luminosity and density evolution. 
Radio sources fall into two principal classes, active galactic nuclei (AGN) and star forming galaxies (SFG). 
The exquisite angular resolution of SKA surveys will allow us to resolve most of the AGNs and thus to 
obtain an extra handle based on morphology. Another possibility to overcome the restrictions 
from evolution is to study the directional fluctuations of differential number counts \citep{Raccanelli,Song}, as 
we do not expect that the properties of radio sources would single out preferred directions in the Universe.  

All SKA forecasts presented in this work assume the baseline design and imaging capabilities as presented in 
\cite{baseline,imaging}. 

\section{Cosmic radio dipole (Early Science, SKA1 \& SKA2)}

The CMB dipole is generally assumed to be due to our peculiar motion and thus defines a cosmic 
reference frame. However, the observation of the dipole in the microwave sky alone does not allow us to tell 
the difference between a motion-induced CMB dipole and dipole contributions from other physical 
phenomena [e.g.~the model in \cite{erickcek:2008jp}]. 

Due to the effects of aberration and Doppler shift, the kinetic dipole must also be present in radio observations 
\citep{Ellis}. Besides the kinetic dipole, we also expect contributions from the large-scale structure and 
from Poisson noise. Such a radio dipole has been looked for in radio source catalogues, such as NVSS 
\citep{Blake:2002gx,Singal,Gibelyou:2012ri,Rubart:2013tx} and WENSS \citep{Rubart:2013tx} and was 
found within large error bars. While the direction of the observed radio dipole is consistent with the 
CMB dipole direction, its amplitude exceeds the theoretical expectations by a factor of a few. SKA will 
enable us to measure the radio dipole with high accuracy and to extract other low-$\ell$ multipole 
moments. Recently, the Planck mission reported a first detection of the effects of aberration and 
Doppler shift at high multipole moments (Planck collaboration \citeyear{Planck2013XXVII}). 
However, this observation is less precise than the reported measurements of the 
radio dipole and allows for a primordial contribution to the CMB dipole of comparable size.

The SKA will allow us to compare $\vec d_{\rm radio}$ to $\vec d_{\rm cmb}$, since SKA will test 
a super-horizon sized volume. Any statistically significant deviation will be exciting, while finding a 
match would put the concordance model on firmer grounds. 

SKA continuum surveys at low frequencies ($< 1\,$GHz) should be ideal to probe the cosmic radio 
dipole already in the Early Science phase for two reasons. First, it is not necessary to cover the full area of the 
$3\pi$ surveys, since a sparse sampling spread out over all of the accessible sky should be sufficient for a 
first estimate. And second, a focus on low frequencies and bright sources will pick primarily AGNs which 
have a much higher mean redshift than the SFG. 

\begin{figure}
\centerline{\includegraphics[height= 4.8 cm]{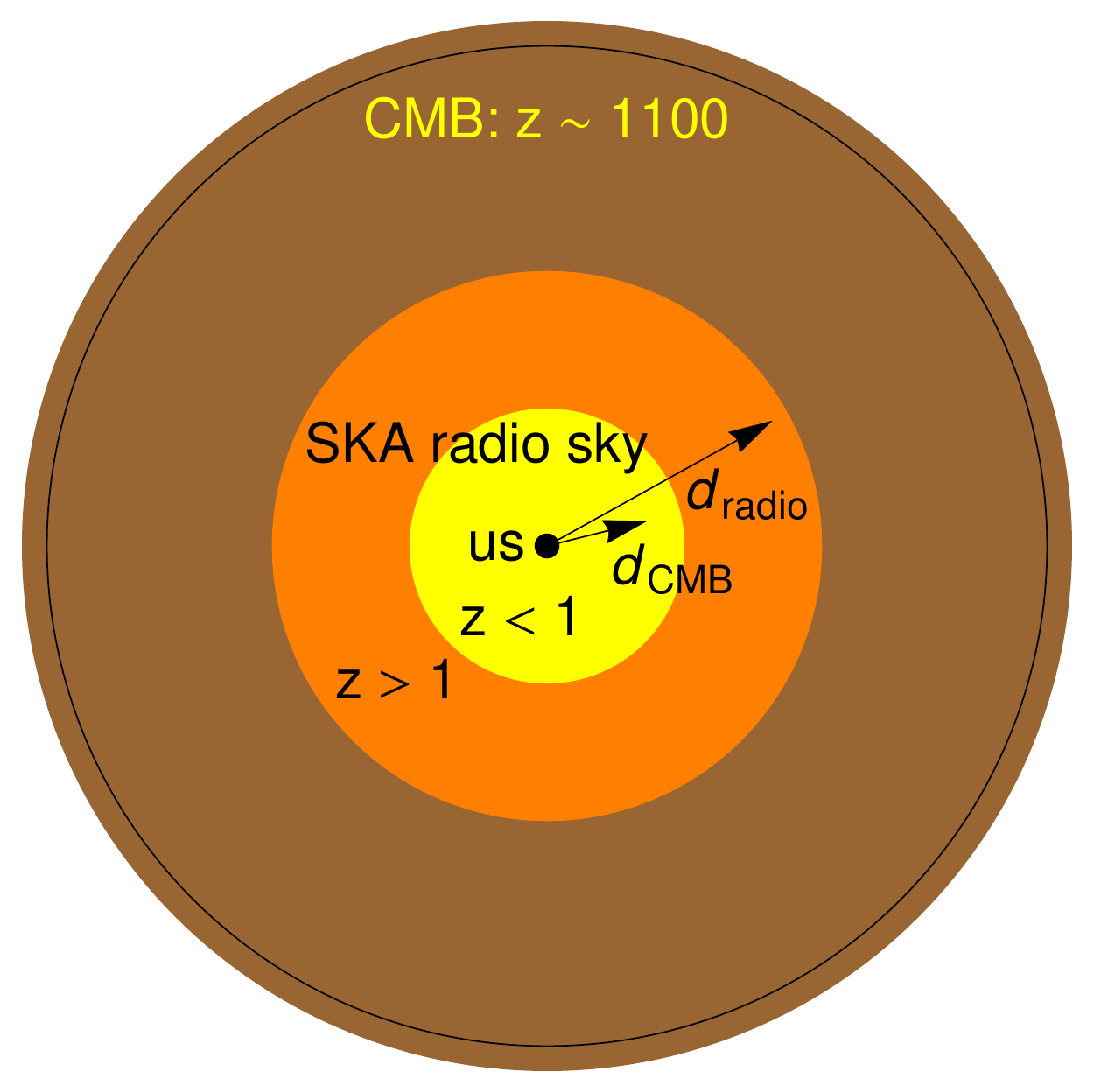} \hspace{1 cm} 
\includegraphics[height = 4.8 cm]{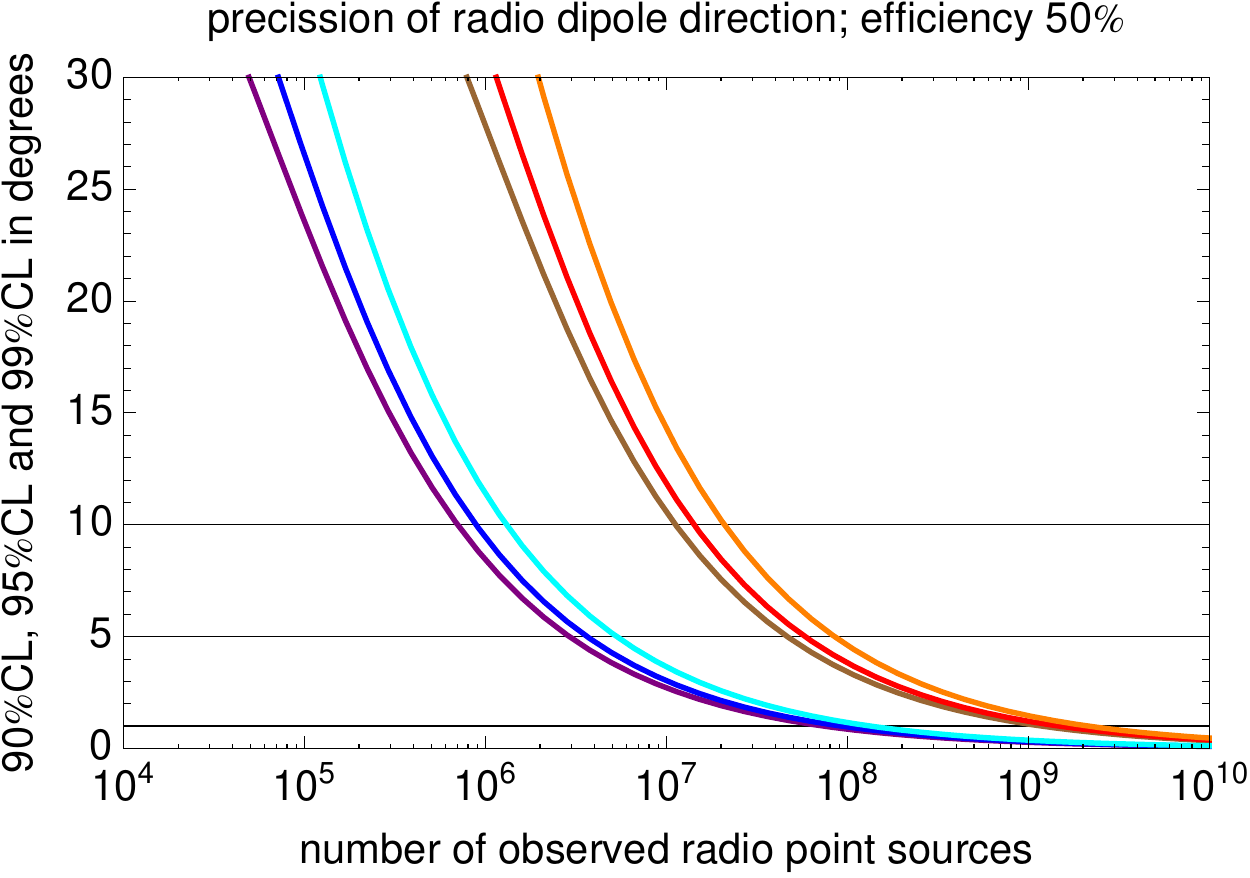}} 
\caption{
\label{fig1}
{\it Left:} 
All-sky (3$\pi$) SKA surveys (yellow and orange) 
will measure the cosmic radio dipole of differential source counts. Selecting AGNs will result in a sample 
with median redshift $z > 1$ (orange) and thus allow us to measure the peculiar velocity of 
the solar system  with respect to the large scale structure on superhorizon scales. These measurements will be 
compared to the CMB dipole and thus test 
for the existence of a bulk flow of our Hubble volume compared to the CMB rest frame.  
{\it Right:} 
Angular accuracy at 90, 95 and 99 \% C.L. of the measurement of the cosmic radio dipole as a function 
of observed point sources. The blue set of curves assumes $d_{\rm radio} = 4 d_{\rm cmb}$, the 
red set assumes  $d_{\rm radio} = d_{\rm cmb}$. It is assumed that only $50\%$ of all detected radio 
sources survive all quality cuts (e.g. masking fields that contain very bright sources). Combined with 
Table 1 we find that SKA Early Science allows detection of a possible deviation from the 
CMB expectation at high significance. SKA1 will constrain the cosmic radio dipole direction 
with an accuracy better than 5 degrees, SKA2 within a degree (at 99\% C.L.).}
\end{figure}

Figure \ref{fig1} illustrates the accuracy that we can hope to achieve for a measurement of the radio dipole 
based on a linear estimator \citep{Crawford, Rubart:2013tx}. Our estimates are based on differential number 
counts from surveys in small and deep fields and simulations \citep{Wilman:2008ew}. Our expectations for 
all-sky continuum surveys are summarized in Table \ref{tab1}.
We find that the cosmic radio dipole can be measured at high statistical 
significance, even taking realistic data cuts into account (e.g. masking the galaxy and very bright extragalactic 
sources, or morphology, spectral index or flux cuts). 

\begin{table}
\begin{tabular}{llll}
SKA $3\pi$ surveys & Early Science & SKA1 & SKA2 \\ 
 (centr.~frequency, ang.~res.)  &  &  & \\
\hline
LOW (151 MHz, 10'') &    $1.0 \times 10^8$ (20 $\mu$Jy) & $2.4 \times 10^8$ (10 $\mu$Jy) & 
$2.2 \times 10^9$ (1 $\mu$Jy) \\
MID/SUR B1 (610 MHz, 1'') &    $7.8 \times 10^7$ (10 $\mu$Jy) & $1.9 \times 10^8$ (5 $\mu$Jy) & 
$1.8 \times 10^9$ (0.5 $\mu$Jy)\\
MID/SUR B2 (1.4 GHz, 0.5'') &  $3.8 \times 10^7$ (10 $\mu$Jy) & $9.7 \times 10^7$  (5 $\mu$Jy) & 
$1.2 \times 10^9$ (0.5 $\mu$Jy) 
\end{tabular}
\caption{Expected total number of radio sources (10 $\sigma$) in various frequency bands and 
survey instruments, assuming the SKA baseline design and the cosmology and differential number 
counts as simulated in \cite{Wilman:2008ew}. In order to match observations at 1.4~GHz, the number of 
SFG has been multiplied by a factor of 2.5 compared to the simulations for all frequency bands. 
The numbers in brackets denote the assumed rms noise levels. \label{tab1}}  
\end{table}

\begin{figure}
\centerline{\includegraphics[height = 7 cm, angle = 270]{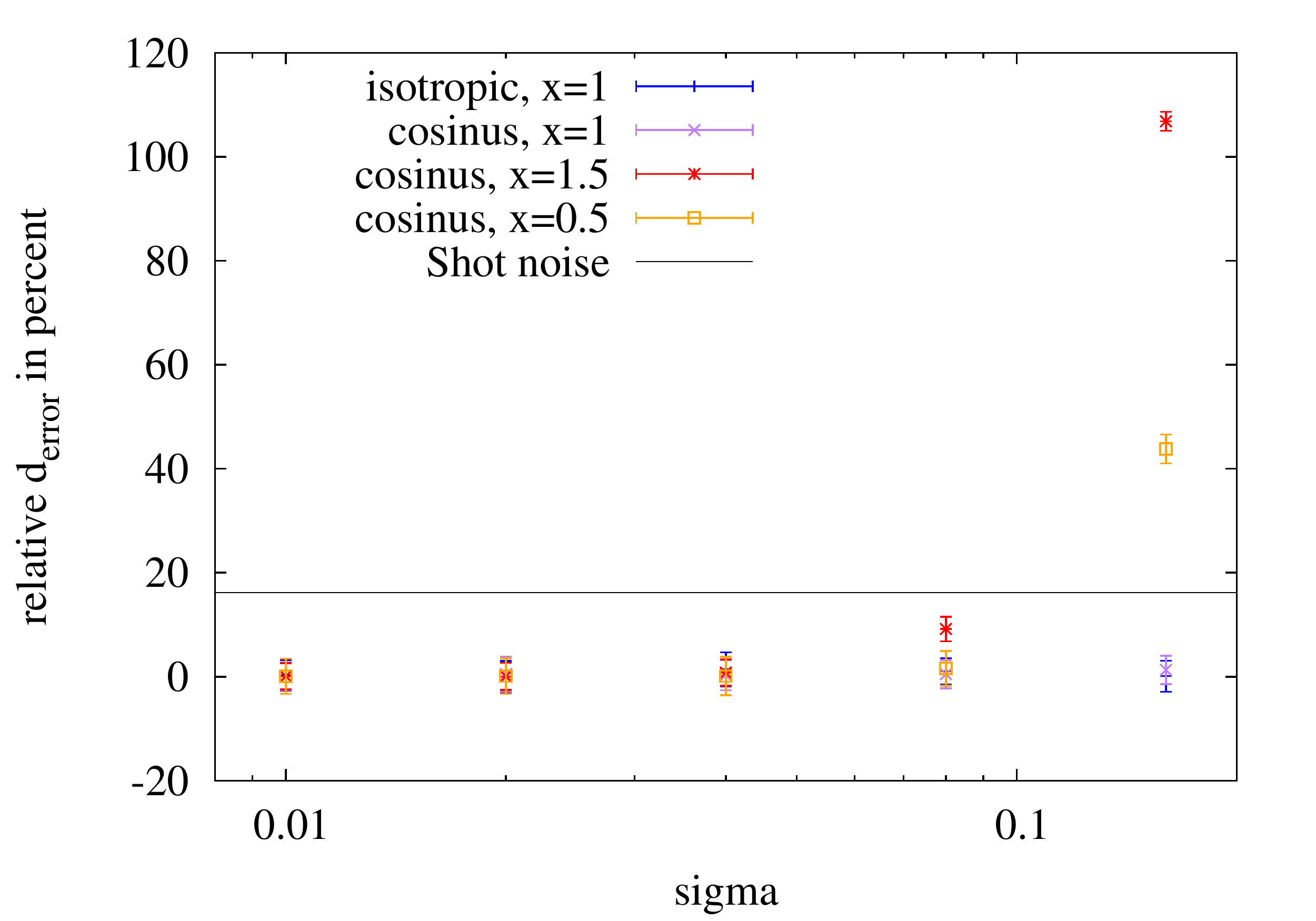}
\includegraphics[height = 7 cm, angle =270]{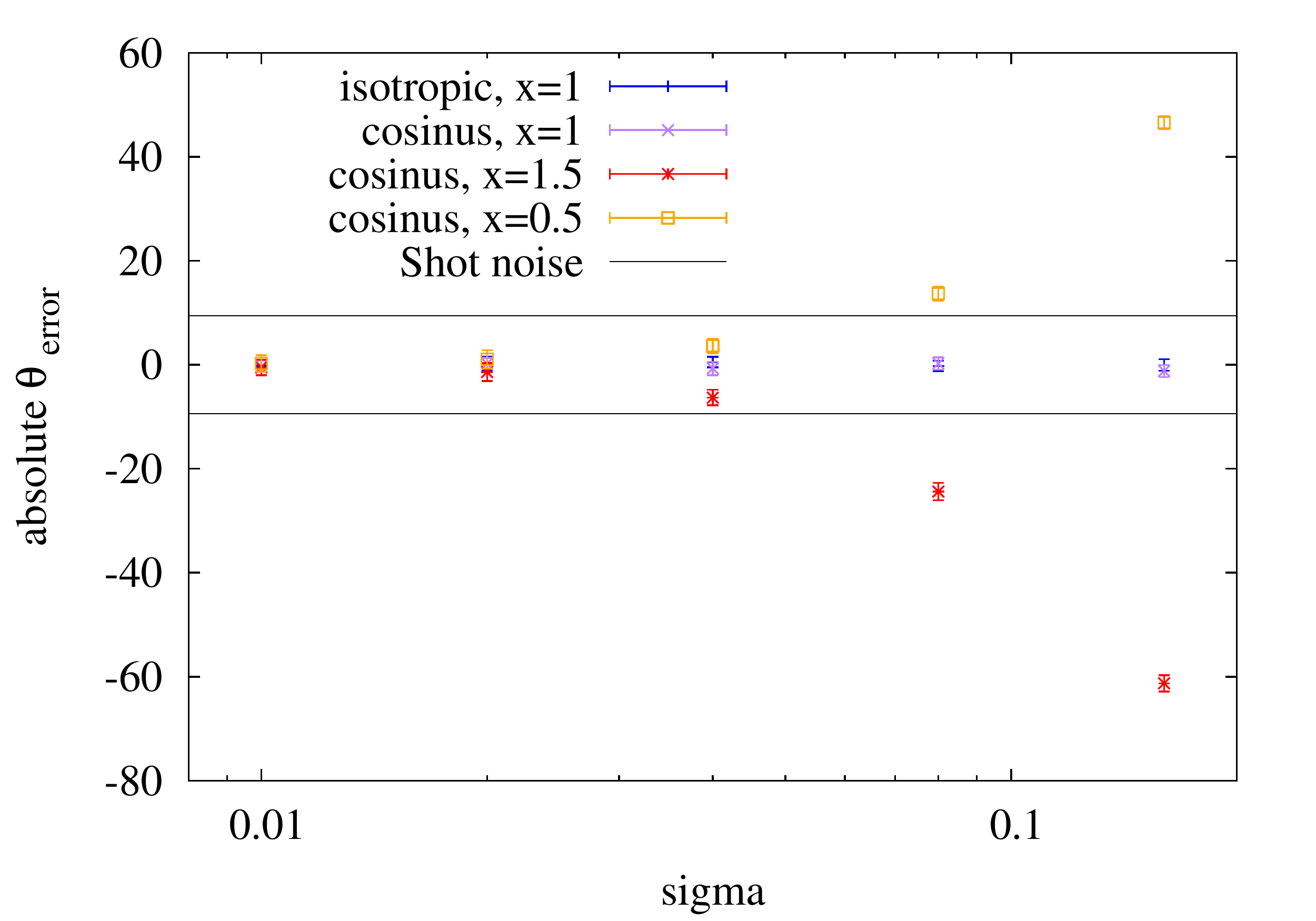}}
\caption{\label{fig2}{\it Left:} Accuracy (in per cent) of the measurement of the dipole amplitude as function of
fractional error on flux density calibration on individual point sources. All points are based on 100 simulations. 
{\it Right:}  Accuracy (in degrees) of the measurement of the dipole direction. The horizontal lines denote 
the error due to shot noise for a dipole estimate based on $10^7$ sources (SKA Early Science).}
\end{figure}

A major concern might be the effect of flux calibration errors on the dipole estimation.  This 
has been studied by means of simulations.  The results of this study are shown in 
figure \ref{fig2}. We assume Gaussian flux density errors with variance $\sigma(\delta) S$, where $S$ 
denotes the expected flux density of a particular source and $\delta$ its declination. 
We consider the isotropic case in which $\sigma(\delta) = \sigma$ is isotropic and a 
declination dependent situation with $\sigma(\delta) = \sigma/\cos(\delta - \delta_*)$, $\delta_*$ being fixed by the latitude of the SKA site 
and $|\delta -\delta_*| < 70$ deg. For two cases we find negligible influence of calibration errors: If the 
flux calibration error is completely isotropic or if the slope $x$ of the number counts 
[$N(>S) \propto S^{-x}$] is equal to one. It turns out that $x=1$ is a special value, where calibration errors 
at the lower flux density limit have no influence on the dipole estimator. We conclude that direction 
dependent calibration effects must not exceed certain limits as shown figure \ref{fig2}.  

\begin{figure}
\centerline{\includegraphics[angle=270, width = 7 cm]{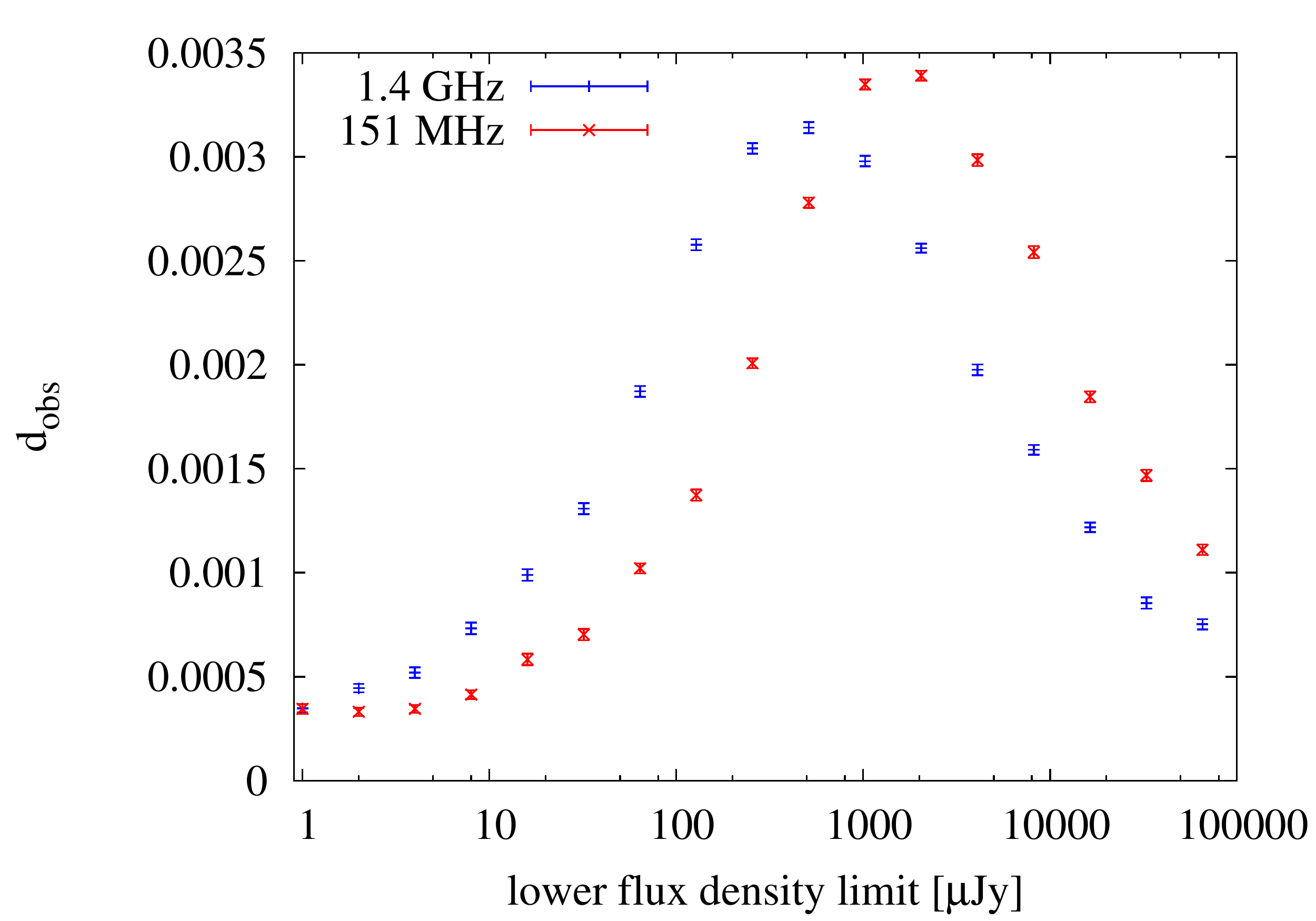}}
\caption{\label{fig3} 
Amplitude of structure dipole due to a local void, affecting the measurements of the cosmic radio dipole 
as a function of lower flux density limit and for two wavebands centered on $151$ MHz and $1.4$ GHz 
(from \cite{Rubart:2014lia}).}
\end{figure}

Another significant contaminant of the kinetic radio dipole is the local structure dipole. We can turn 
a disadvantage of continuum surveys, namely that we observe several source populations, into an 
advantage as follows: The lower mean redshift of SFGs compared to AGNs allows us to change the mean 
depth of the survey by scanning different fluxes density limits and frequencies. This in fact allows for 
a tomographic survey of the radio dipole. For the example of a huge ($\sim 100$~Mpc) local void 
this was studied recently \citep{Rubart:2014lia}. Figure \ref{fig3} illustrates this effect.
 
\section{Large angular scales (SKA1 \& SKA2)} 

It is not obvious that the isotropic distribution of light implies also the isotropy of
space-time itself. The vanishing of the quadrupole and octopole moments of the 
CMB would imply the isotropy of space-time along the world line of the observer \citep{Maartens:2011yx}. 
While those low-$\ell$ multipoles are small compared to the monopole and dipole, they do not vanish 
exactly. We thus we can at best speak about an almost isotropic Universe. 
The radio sky offers another independent probe at $z > 1$ and at the largest angular scales.

Recent work has revealed the existence of CMB ``anomalies" [for a review,
see \cite{Copi_AdvAstro}]. In brief, the angular correlation function in the
WMAP and Planck temperature maps vanishes on scales larger than 60 degrees,
contrary to theoretical expectation; moreover, the CMB quadrupole and octopole
anisotropy patterns are  aligned both mutually and with respect
to the Solar System geometry. These anomalies have been widely studied and
discussed, but their origin remains unexplained. 

SKA will provide a deep and wide large-scale structure dataset that will 
enable separating the effects of the early and late universe on the observed
CMB anisotropy. For example, the SKA data could be used to reconstruct the
late-time contribution to the CMB anisotropy via the integrated Sachs-Wolfe
effect, and thus provide information about the temporal evolution of the CMB
anomalies.

\subsection{Low-$\ell$ multipole moments} 

\begin{figure}
\centerline{\includegraphics[height = 4.5 cm]{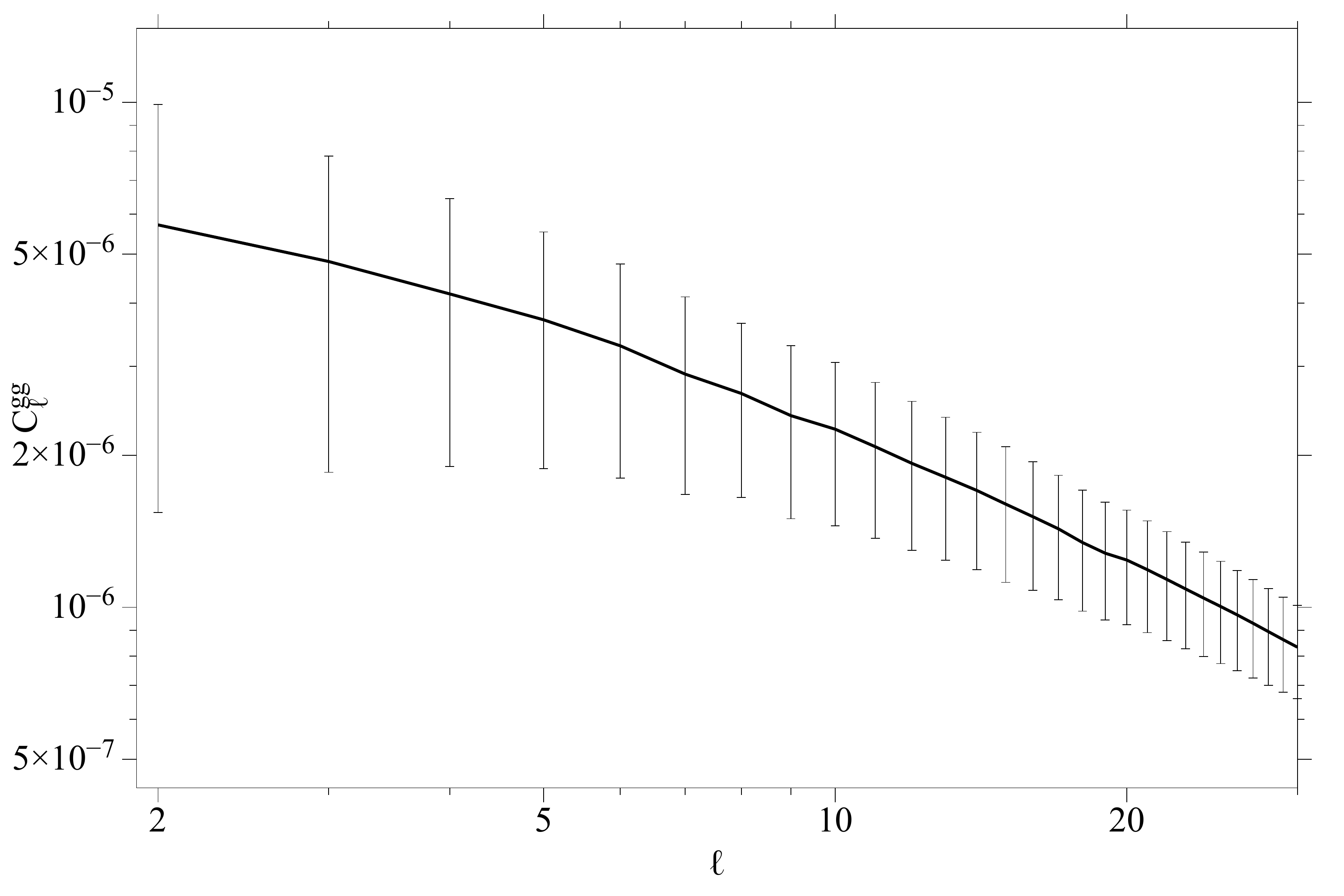}
\includegraphics[height = 4.5cm]{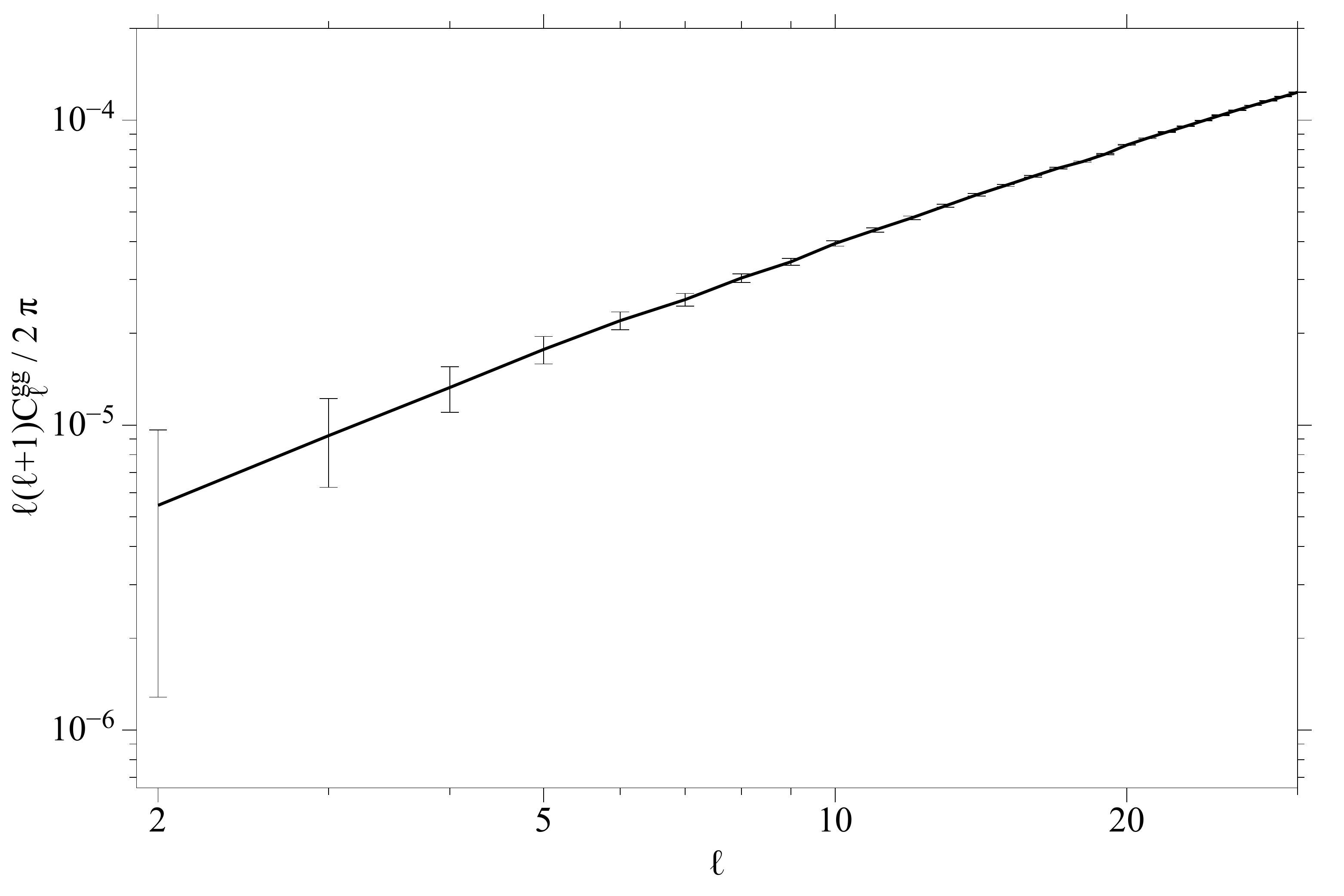}}
\caption{\label{fig4} 
{\it Left:} Low-$\ell$ multipoles of the angular power spectrum as expected for a SKA1 continuum 
survey. {\it Right:} The corresponding band power. 
The errors contain shot noise and cosmic variance. The figures illustrate that statistically significant 
measurements of multipole moments can be expected.}
\end{figure} 

The analysis of low-$\ell$ multipoles from SKA continuum surveys can benefit from 
the methods developed for the study of the CMB.  Missing sky area is always a problem 
for low-$\ell$ mode measurements. For CMB
studies, many methods were proposed to deal with a mask of missing data for both 
power spectrum estimation and phase recovery. For the power spectrum, one of the most used 
methods is MASTER \citep{Hivon2002}. It consists in first building a matrix
which captures the coupling between the modes induced by the mask, and then 
inverting this matrix. In Figure 4 we plot the angular power spectrum of SKA galaxies 
for low-$\ell$ mulitpoles with error bars corresponding to a SKA1 continuum survey.

For phase recovery, or more generally for large scale map reconstruction,
many methods have been proposed based on Wiener filtering,
$l_2$ or $l_1$ norm regularization, constraint realizations or diffusion
[see \cite{Starck2013} and references therein].
Based on these new methodological idea, Planck data were analyzed with a
mask removing 27\% of the sky (Planck collaboration \citeyear{Planck2013XXIII}; 
Rassat et al. \citeyear{Rassat2014}).
For a given observed sky area, the shape of the mask will also be
important. The importance of random sampling is also described
in  \cite{Paykari2013}, and many small missing parts, randomly distributed,
will always be much better for large scale studies than a compact big
missing part. 

As for the dipole, SKA will tremendously improve the precision and quality of low-$\ell$ multipole moments 
and thus allow us to probe statistical isotropy, scale invariance and gaussianity.  

\subsection{Angular 2-point correlation function} 

The angular two point correlation is a powerful tool to measure the projected large-scale structure distribution of the Universe. 
It allows us to probe certain fundamental assumptions like scale-invariance of the primordial 
perturbations, Gaussianity and the isotropy of the Universe 
(by comparing two-point correlations on sub-samples of the observed sky).  

The two-point correlation at large angular scales contains many interesting aspects: 
Firstly, the matter fluctuations at large scales are in the linear regime.
Secondly, general relativistic effects and cosmological evolution prefer
large scales. AGNs are very good candidates to probe this, since they are 
isotropically distributed on the sky and most of them have a significant ($z > 1$) 
cosmological distance. In order to accurately investigate
ultra-large scale correlations, the theoretical frame work of differential number counts
based on general relativity will be needed  
\citep{Maartens:2012rh,Raccanelli:2013dza,Raccanelli:2013gja,Song}. 

In contrast to galaxy redshift surveys within the local Universe ($z \ll 1$), all linear order relativistic corrections,
which include the Doppler effect, lensing, and generalized Sachs-Wolfe contributions, are of relevance. 
The spread of luminosities of radio sources in continuum surveys washes out much of the clustering
signal, and the general relativistic corrections are also suppressed. However, 
SKA HI surveys in which the source redshifts will be known will resolve these effects.

With the assistance of Lyman alpha data, one can model the luminosity
function and evolution. With the SKA morphology data we expect to be able to identify
different type of sources. This will allow us to study cross correlations between 
star forming galaxies and AGNs. We could also cross correlate with the CMB
and different types of radio sources, which have different redshift distributions.

These aspects are treated in more detail in other contributions to this volume (Camera et al.~\citeyear{Camera};
Bull et al.~\citeyear{BAO}; Raccanelli et al.~\citeyear{RSD}).
Let us just stress here the importance of re-establishing the almost scale-invariant power spectrum 
at superhorizon scales at $z \sim 1$, which will be possible by means of SKA all-sky surveys.

\section{Copernican Principle and homogeneity (SKA1 \& SKA2)}

The Copernican Principle is the assumption that we are not distinguished observers in the Universe. 
If we observe an isotropic cosmos, then distant observers should also see a similarly isotropic cosmos. 
This implies that the Universe satisfies the Cosmological Principle and is homogeneous on large scales.  
A violation of homogeneity in principle offers an alternative explanation to the acceleration of the Universe 
\citep{Celerier:1999hp}, but simple inhomogeneous models without dark energy are incompatible with 
current data~\citep{Bull:2011wi}. However, radial homogeneity is only weakly constrained in 
$\Lambda$CDM~\citep{Valkenburg:2012td}. Any deviations would imply a radical change to the standard 
model and scale-invariant initial conditions, making it a vital constraint on the standard model.

SKA HI intensity mapping 
on super-Hubble scales offers powerful new ways to test homogeneity. By comparing the radial and 
transverse scale of baryon acoustic oscillations we can test isotropy of the expansion rate around distant 
observers~\citep{Maartens:2011yx,February:2012fp,Clarkson:2012bg}.  This places direct constraints on 
radial inhomogeneity about us, when redshift-space distortions, lensing and other large-scale GR effects 
are accounted for. 
Anisotropic expansion rates act on the sound horizon at decoupling so that by redshift $z$ it has evolved 
into an ellipsoid with semi-axes
 \begin{equation}
 L_\|(z) ={\delta z(z) \over (1+z)H_\|(z)}\,,~~~L_\perp(z)   =d_A(z) \delta\theta(z), 
 \end{equation}
given the observed radial and angular scales $\delta z(z), \delta\theta(z)$.
$L_\|(z)=L_\perp(z) $ in a homogeneous universe. In an inhomogeneous universe $d_A(z)$ depends on 
the transverse Hubble rate along the line of sight, which will be different from the radial Hubble rate $H_\|(z)$, 
providing a test of homogeneity.
  
When combined with accurate distance data from SNIa, consistency relations can be used to check deviations 
from homogeneity in a completely model independent way~\citep{Clarkson:2007pz}. In a homogeneous universe, 
irrespective of dark energy or theory of gravity, the Hubble rate $h(z)=H(z)/H_0$ and dimensionless comoving 
distance $D(z)=(1+z)H_0d_A(z)$ satisfy $({}^\prime=d/dz)$
\begin{equation}\label{C(z)}
\mathscr{C}(z)=1+h^2\left(DD''-D'^2\right)+hh'DD'=0\ ,
\end{equation}
so that $\mathscr{C}(z)\neq0$ implies violation of the Copernican Principle. We expect that SKA1 will be able to 
constrain $\mathscr{C}(z)$ to $0 \pm 0.05$ for $z < 1.5$, based on a naive error propagation from 
\cite{Bull:2014rha}. A more careful forecast has yet to be done.  
Direct constraints on radial inhomogeneity can be given combining with all available data sets which will 
significantly improve current constraints which are much weaker than those for 
isotropy~\citep{Valkenburg:2012td}. 

Finally, the Copernican Principle allows for the possibility of a 
fractal universe, but this is not predicted by the concordance model~-- which predicts a fractal dimension of 
3 on large scales~-- any deviations would imply new physics. It is therefore important to measure the fractal 
dimension of the distribution of radio sources at superhorizon scales. Such a test has been performed using 
the SDSS and the WiggleZ surveys, finding an approach to a three-dimensional distribution at  
$\sim100$~Mpc scales \citep{SDSSfractal, WiggleZfractal}.  A dramatic improvement will be possible 
based on SKA HI threshold surveys.

\section{Summary} 

The Cosmological Principle provides the foundation for modern cosmology,
and our understanding of the evolution of the Universe as well as all
parameter constraints from the CMB, supernovae or large scale structure rely
on this assumption. Testing the Cosmological Principle is thus of
fundamental importance for cosmology generally as well as for the
cosmological interpretation of the SKA data itself. As this chapter shows,
SKA will be able to greatly increase our confidence that our cosmological
framework makes sense (or lead to a scientific revolution if not).

We argue that SKA all-sky surveys will allow us to measure the cosmic radio dipole almost as precisely as the 
CMB dipole. SKA1 will constrain the cosmic radio dipole direction with an accuracy better than 5 degrees, 
SKA2 within a degree (at 99\% C.L.). This measurement could finally firmly establish or refute 
the commonly adopted assumption that the CMB and the overall LSS frames agree, and will have impact on 
a variety of cosmological observations, from the local measurement of $H_0$ to the calibration of 
CMB experiments. A tomography of the cosmic radio dipole might reveal a detailed understanding of 
local LSS. 

In addition, studying the large-angular scales in SKA continuum and HI surveys might help resolve the 
puzzle of CMB anomalies and test the cosmological principle, including tests of statistical homogeneity. 
Further large-scale structure issues, especially non-Gaussianity and relativistic corrections, are discussed 
in \cite{Camera}. 

The ideas presented in this work only provide a flavor of SKA's potential to answer 
fundamental cosmological questions. Some of those ideas can already be tested by means of the SKA 
pathfinder experiments ASKAP, MeerKAT and LOFAR, but they cannot compete with SKA's survey speed 
and sensitivity. Thus SKA will be a unprecedented discovery and precision machine for modern 
cosmology. 

\section*{Acknowledgments}

DJS, SC and MR are supported by the Deutsche Forschungsgemeinschaft by means of the Research Training Group 1620 ``Models of Gravity''. MR acknowleges financial support from the Friedrich Ebert-Stiftung. 
AR is supported by the Templeton Foundation. Part of the research described in this work was carried 
out at the Jet Propulsion Laboratory, California Institute of Technology, under a contract with the National Aeronautics and Space Administration.

\bibliographystyle{apj}  
\bibliography{AASKA14_32} 

\end{document}